\documentclass[amsmath,superscriptaddress,amssymb,aps,twocolumn,prl]{revtex4}
\usepackage{graphicx}
\usepackage{soul}
\usepackage[colorlinks=true,citecolor=blue,linkcolor=blue]{hyperref}

\usepackage[usenames]{color}

\newcommand{\ket}[1]{\left\vert #1 \right\rangle}
\newcommand{\bra}[1]{\left\langle #1 \right\vert}
\newcommand{\ketbra}[2]{\ket{ #1}\bra{ #2} }
\newcommand{\bla}[1]{\left( #1 \right)}
\newcommand{\blb}[1]{\left[ #1 \right]}

\def \ket#1{|#1\rangle}
\def \bra#1{\langle#1|}

\def \be{\begin{equation}}
\def \ee{\end{equation}}
\def \ba{\begin{array}}
\def \ea{\end{array}}
\def \bea{\begin{eqnarray}}
\def \eea{\end{eqnarray}}

\renewcommand{\phi}{\varphi}

\begin{document}

\title{Chemical compass model for avian magnetoreception as a quantum coherent device}
\author{Jianming Cai}

\affiliation{Institut f\"{u}r Theoretische Physik, Albert-Einstein Allee 11, Universit\"{a}t Ulm, 89069 Ulm, Germany}
\affiliation{Center for Integrated Quantum Science and Technology, Universit\"{a}t Ulm, 89069 Ulm, Germany}
\author{Martin B. Plenio}
\affiliation{Institut f\"{u}r Theoretische Physik, Albert-Einstein Allee 11, Universit\"{a}t Ulm, 89069 Ulm, Germany}
\affiliation{Center for Integrated Quantum Science and Technology, Universit\"{a}t Ulm, 89069 Ulm, Germany}

\date{\today}

\begin{abstract}
It is known that more than 50 species use the Earth's magnetic field for orientation and navigation. Intensive studies particularly behavior experiments with birds, provide support for a chemical compass based on magnetically sensitive free radical reactions as a source of this sense. However, the fundamental question of how quantum coherence plays an essential role in such a chemical compass model of avian magnetoreception yet remains controversial. Here, we show that the essence of the chemical compass model can be understood in analogy to a quantum interferometer exploiting global quantum coherence rather than any subsystem coherence. Within the framework of quantum metrology, we quantify global quantum coherence and correlate it with the function of chemical magnetoreception. Our results allow us to understand and predict how various factors can affect the performance of a chemical compass from the unique perspective of quantum coherence assisted metrology. This represents a crucial step to affirm a direct connection between quantum coherence and the function of a chemical compass. 
\end{abstract}

\maketitle

{\it Introduction.---} Despite the growing interest from chemists, biologists and more recently researchers from quantum physics and quantum information, there remain to date only a handful of biological phenomena that are suspected or proven to rely on quantum effects. These include  important biological processes such as light harvesting \cite{Engel07,Collini10,LHCT1,Plenio08,Scho11,Chin13}, human sense of smell \cite{SME1,SME2,SME3}, and avian magnetoreception \cite{Sch78,Ritz00,Sol07,Rodgers09,Ritz10,Ritz09,Ritz11,Hore10,Maeda08,Kom09,Cai10,Gauger11,Cai12,Band12,Hogben12,Sun12,Guer2013}. The effects of weak magnetic field in nature have been observed for a long time \cite{Joh05}, ranging from the growth of plants to the remarkable orientation and navigation abilities of animals such as birds and insects. The radical pair mechanism based on anisotropic hyperfine interactions, as a leading theory to explain avian magnetoreception \cite{Werner77}, suggests that the avian compass relies on magnetically sensitive radical pairs formed by photoinduced electron transfer reactions \cite{Sch78,Ritz00,Sol07,Ritz10,Ritz09,Rodgers09}. The cryptochromes in the retina of migratory birds provides a potential physiological implementation of such a mechanism \cite{Mour04,Weaver00nature,Hore03,Solo07,Ahmad07,Ritz00,Zap09,Maeda12,Sol10,Sol12,Lied10}. The observations from the behavior experiments \cite{WiltsRev05,Phil92,Ritz04nature} with birds provide corroborating evidence for the idea that the chemical compass mechanism is involved in avian magnetoreception. 

In the last few years, the interest in avian magnetoreception has quickly extended from chemists and biologists to quantum physicists \cite{Ritz11,Hore10,Kom09,Cai10,Cai12,Sun12,Gauger11,Band12,Hogben12}. In the chemical compass model, a radical pair is born in spin singlet or triple states \cite{Sch78,Steiner89,Ritz00}. The subsequent dynamics is composed of a quantum coherent interaction with nearby nuclei and the external magnetic field. The former is usually considered as a noise process that suppresses quantum coherence. In previous consideration emphasis was placed on the electronic coherence properties and it remained unclear how this coherence was directly exploited in the function of a chemical compass \cite{Cai10,Cai12,Gauger11,Band12,Hogben12}. For example, a chemical compass can still show good sensitivity even though it exhibits negligible electronic coherence \cite{Cai10,Hogben12}. It is not clear how to establish a quantitative relation between coherence and compass sensitivity.

\begin{figure}[t]
\begin{center}
\hspace{-0.3cm}
\includegraphics[width=8.5cm]{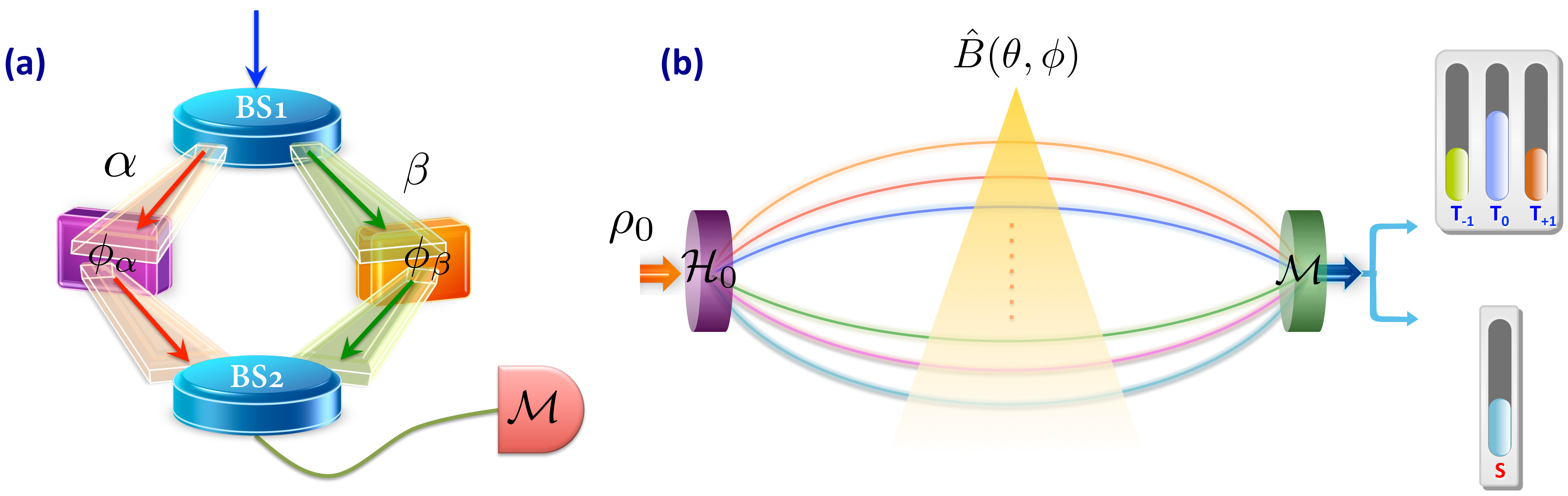}
\end{center}
\caption{Quantum interferometer model of a chemical compass. (a) A quantum system is prepared ($\mbox{BS}_1$) in a coherent superposition of two quantum states $\ket{\alpha}$ and $\ket{\beta}$, which gains different dynamical phases dependent on an unknown physical parameter, the information of which is revealed by the measurement after the interference ($\mbox{BS}_2$). (b) A chemical compass is viewed as a quantum interferometer, the hyperfine coupling Hamiltonian $\mathcal{H}_0$ recasts the state $\rho_0$ of the combined system of radical pair and nuclei into a superposition of its eigenstates. The magnetic field induces the change in coherence phases arising from the dynamic phases of individual eigenstate evolution. The spin-dependent chemical reaction channels lead to the interferometric observable which provides the information about the direction of the magnetic field.}\label{fig:INF}
\end{figure}

In the present work,we address these fundamental questions and reveal the role of quantum coherence in chemical magnetoreception by appreciating that the quantum dynamics of a chemical compass giving rise to the magnetic sensitivity is akin to a quantum interferometer in quantum metrology \cite{Wine94,Budker07,Caves07}. Such a perspective allows us to establish a quantitative connection between compass sensitivity and the global coherence with respect to the eigenbasis of the hyperfine Hamiltonian of the combined system of radical pair electrons and the surrounding nuclei. We verify that the concept of global coherence is indeed essential for the function of a chemical compass by introducing an appropriate quantification of coherence which has an operational meaning in the context of chemical compass. The results are shown to be valid for general radical pair molecules, and we further explicitly demonstrate the idea for the radical pair inspired by the flavin adenine dinucleotide (FADH$\cdot$) formed photochemically in cryptochromes (which is a very probable candidate for avian magnetoreception \cite{Ritz00,Hore03,Maeda12,Sol10,Sol12}) in more detail.

{\it Quantum interferometer model of chemical compass.---} In quantum metrology, a system, e.g., a photon or a spin, is initially prepared into the following state
\be
\rho_0=\ketbra{\Psi_0}{\Psi_0}, \quad \mbox{with  } \quad \ket{\Psi_0}=\gamma_{\alpha}\ket{\alpha}+\gamma_{\beta}\ket{\beta},
 \ee
which is written in the basis of $\{\ket{\alpha},\ket{\beta}\}$ (determined by the interferometric element BS1, see Fig.\ref{fig:INF}). A Hamiltonian $H(\lambda)=\frac{\lambda}{2}\bla{\ketbra{\alpha}{\alpha}-\ketbra{\beta}{\beta}}$ parametrized by a single unknown parameter $\lambda$ results in an evolution that changes the relative phase between the two states $\ket{\alpha}$ and $\ket{\beta}$ to $\Phi_\lambda=\phi_{\alpha}-\phi_{\beta}$, see Fig.\ref{fig:INF}(a), and leads to the state 
\be
\rho_0\rightarrow \rho_{\Phi_{\lambda}}=\left( \begin{array}{cc}
|\gamma_{\alpha}|^2& \gamma_{\alpha} \gamma^{*} _{\beta} e^{i\Phi_{\lambda}}\\
\gamma_{\alpha} ^{*}\gamma _{\beta} e^{-i\Phi_{\lambda}} & |\gamma_{\beta}|^2 \end{array} \right).
 \ee
The measurement of the observable $\mathcal{M}=\ket{\Psi_0}\bra{\Psi_0}$, gives the interference (see BS2 in Fig.\ref{fig:INF}) outcome as
\be
m_{\lambda}(\rho_0,\Phi_\lambda)\equiv\mbox{Tr} \bla{\rho_{\Phi_\lambda} \mathcal{M}} = 1-2|\gamma_{\alpha}|^2|\gamma_{\beta}|^2\bla{1-\cos{\Phi_\lambda}}.\label{eq:mainf}
\ee
The precision of parameter estimation with such a basic quantum interferometer is determined by the contrast of the interference fringe as follows 
\be
\mbox{D}_{\vec{\gamma}}(\rho_0)\equiv 2|\gamma_{\alpha}|^2|\gamma_{\beta}|^2 \blb{\max_{\Phi_{\lambda}}\cos{\Phi_\lambda}-\min_{\Phi_{\lambda}}\cos{\Phi_\lambda}}.
\ee
The role of coherence in such a simple interferometric setup can be quantified by
\be
\mathcal{C}\equiv |m_{\lambda}(\rho_0^c,\Phi_{\lambda})|_{\Phi_{\lambda}=0}=2 |\gamma_{\alpha}|^2|\gamma_{\beta}|^2,
\label{eq:mcqm}
\ee
with
\be
\rho_0^c = \left( \begin{array}{cc}
0& \gamma_{\alpha} \gamma^{*} _{\beta}\\
\gamma_{\alpha} ^{*}\gamma _{\beta} & 0 \end{array} \right),
\ee
which characterizes how coherent the quantum interferometer is. By comparing Eq.(4) and (5), it is clearly seen that how coherence plays its role and determines the precision of parameter estimation in quantum interferometric metrology. 

\begin{figure}[t]
\begin{center}
\hspace{-0.35cm}
\includegraphics[width=4.45cm]{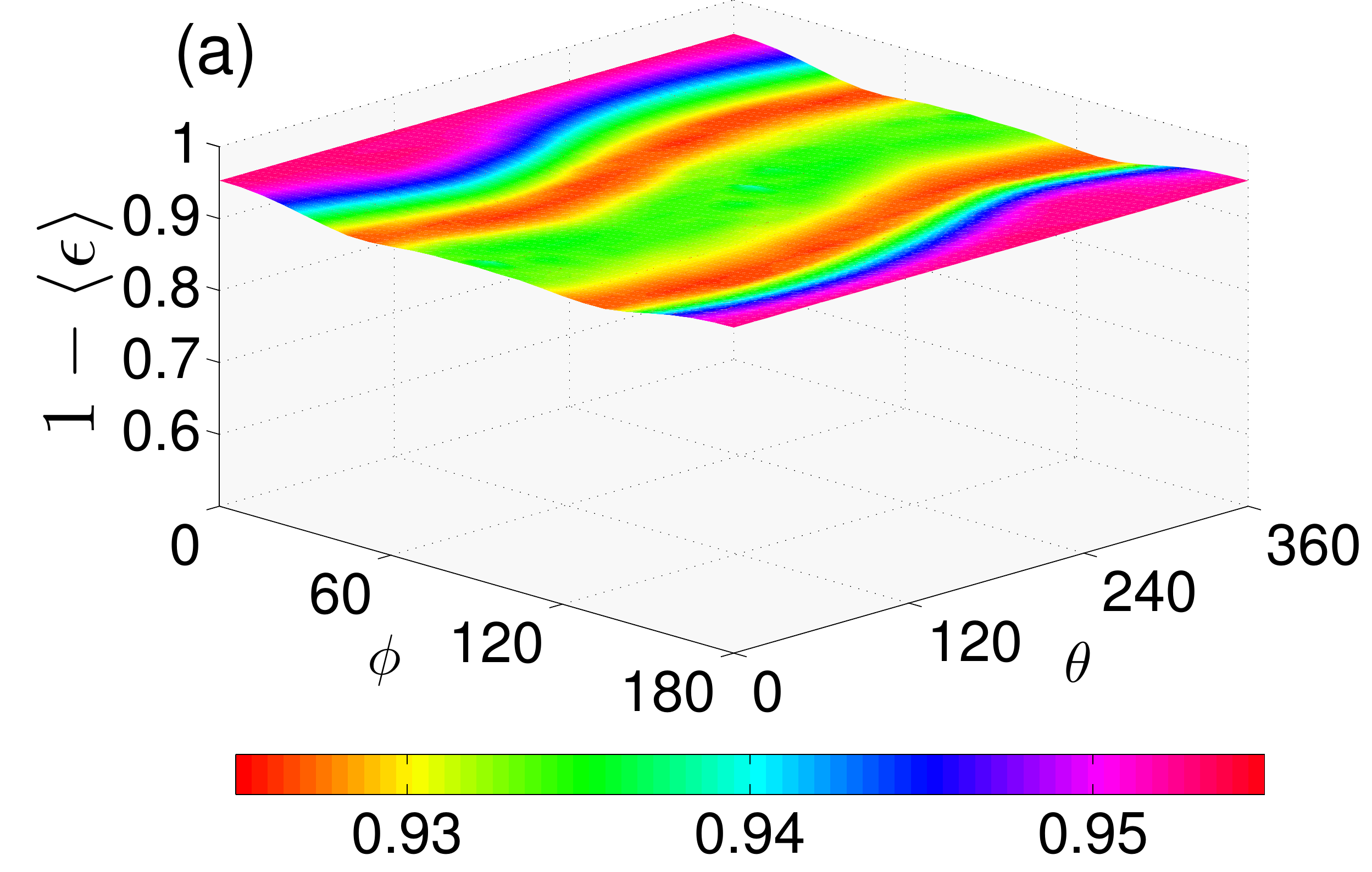}
\hspace{-0.2cm}
\includegraphics[width=4.45cm]{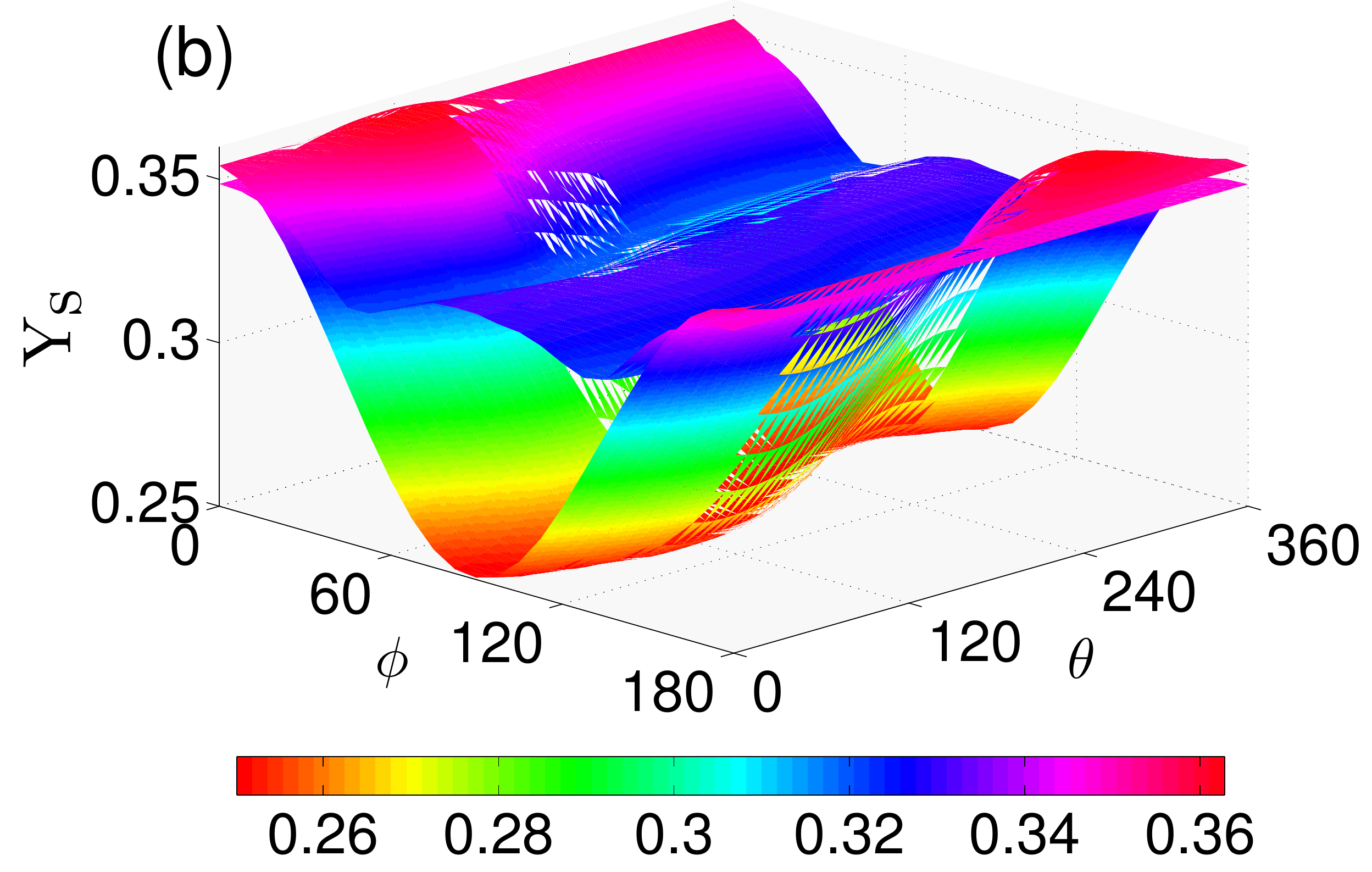}
\end{center}
\caption{Effective evolution of a radical pair (in which one radical contains three $^1$H nuclei and two $^{14}$N nuclei as FADH$\cdot$). (a) The change of the absolute values of the (coherent) density elements $\langle\epsilon(t) \rangle $ averaging over time, with $\epsilon(t)=[\sum_{m\neq n} (r_{mn}^t-r_{mn})^2]^{1/2}$, see Eq.(\ref{eq:recast}-\ref{eq:totalevolution}), for different magnetic field directions $(\theta,\phi)$. (b) The singlet yield $\mbox{Y}_{\mbox{s}}$ as a function of the magnetic field direction $(\theta,\phi)$ including (lower) {\it vs.} excluding (upper) the coherence phase changes $\phi_{mn}^{\hat{B}}(t)$ resulting from the Earth's magnetic field, see Eq.(\ref{eq:totalevolution}). The reaction rates are $k_S=k_T=0.5 \mu s^{-1}$, the Earth's magnetic field is $b=50\mu T$.}\label{fig:FID}
\end{figure}

In the standard model of the chemical compass, the magnetically sensitive radical pairs formed by photoinduced intramolecular electron transfer reactions interact with a few nearby nuclei via hyperfine couplings, as described by the following Hamiltonian
\be
\mathcal{H}_0=\sum_{k=\mathcal{D,A}}\sum_{j}\vec{\mathbf{s}}_{k}\cdot \hat{\mathbf{T}}_{k_{j}}\cdot \vec{\mathbf{I}}_{k_{j}}  
\ee
where $\vec{\mathbf{I}}_{k_{j}}$ and $\vec{\mathbf{s}}_{k}$ are the nuclear and electron spin operators respectively, $\hat{\mathbf{T}}_{k_{j}}$ denotes the hyperfine coupling tensors \cite{Steiner89}. For simplicity, we neglect the dipole-dipole and exchange interactions between two radicals, which is valid when the radical-radical distance is sufficiently large or they cancel with each other \cite{Efi08}. The joint state $\rho_0$ of the radical pair and nuclear spins, written in the eigenbasis $\{\ket{\nu_m}\}$ of the hyperfine interaction Hamiltonian $\mathcal{H}_0$, can be expressed as $\rho_0 = \sum_{m,n} r_{mn} \ket{\nu_m}\bra{\nu_n}, \quad \mbox{with} \quad r_{mn}=\bra {\nu_m}\rho_0\ket{\nu_n}$. In the absence of a magnetic field, the state evolution solely under the hyperfine interaction Hamiltonian $\mathcal{H}_0$ is 
\be
\rho^{\mathcal{H}_0}(t) = \sum_{m,n} r_{mn}  e^{-i\phi_{mn}^{0}(t)}\ket{\nu_m}\bra{\nu_n}.
\label{eq:recast}
\ee
For a typical radical pair molecule with a few nuclei, the leading order effect of the Earth's magnetic field $\mathrm{\hat{B}}=-g\mu_b \vec{b}\cdot (\vec{\mathbf{s}}_{\mathcal{D}}+\vec{\mathbf{s}}_{\mathcal{A}})$ (with $|\vec{b}|=50\mu \mbox{T}$, which is much smaller than the hyperfine interaction strength) is to introduce an additional magnetic field dependent phase on the system coherence according to the perturbation theory, namely 
\be
\rho^{\mathcal{H}_0+\mathrm{\hat{B}}(\theta,\phi)}(t) \approx \sum_{m,n} r_{mn}^t  e^{-i\blb{\phi_{mn}^{0}(t)+\phi_{mn}^{\hat{B}}(t)}}\ket{\nu_m}\bra{\nu_n},\label{eq:totalevolution}
\ee
where $\rho^{\mathcal{H}_0+\mathrm{\hat{B}}(\theta,\phi)}(t)$ represents the exact state evolution under the total Hamiltonian  including both the hyperfine coupling and the Earth's magnetic field. The effect of the Earth's magnetic field might also change the absolute values of the coherent off-diagonal density matrix elements $r_{mn}$ ($m\neq n$), which can be quantified by $\epsilon(t)=[\sum_{m\neq n} (r_{mn}^t-r_{mn})^2]^{1/2}$. In Fig.\ref{fig:FID}(a), we plot $\epsilon$ averaging over time for the radical pair inspired by the flavin adenine dinucleotide formed photochemically in cryptochromes as an example, which shows that the change of the absolute values of the density matrix elements is very small.

The spin-dependent reactions in a chemical compass, i.e. the singlet and triplet radical pairs will undergo different chemical reaction paths and thereby lead to different chemical consequences \cite{Ritz00}, that witness the magnetic field effect on the radical pair dynamics by the magnetic anisotropy of reaction yield. For simplicity, we consider the scenarios where the reaction rates of the singlet and the triplet radicals are identical, i.e. $k_S=k_T\equiv k$ (see \cite{SI} for the generalization to more general cases). Treating the system dynamics with the conventional Haberkorn approach \cite{Steiner89}, the singlet yield can be formulated as 
\be
\mbox{Y}_{\mbox{s}}(\mathcal{H}_0,\rho_0,\theta,\phi) = \int \mbox{Tr}\blb{\rho^{\mathcal{H}_0+\mathrm{\hat{B}}(\theta,\phi)}(t)  \mathcal{M} }dt
\label{singletyield}
\ee
with the observable $\mathcal{M}=k  e^{-kt} ( \ket{\mathcal{S}}\bra{\mathcal{S}}\otimes_{k_j} \mathbb{I}_{d_{k_j}})$, where $\ket{\mathcal{S}}=\sqrt{\frac{1}{2}}\bla{\ket{\alpha\beta}-\ket{\beta\alpha}}$. The singlet yield in Eq.(\ref{singletyield}) can be viewed as a continuous generalization of the outcome in a quantum interferometer as $m_{\lambda}(\rho_0,\Phi_{\lambda})$, see Eq.(\ref{eq:mainf}). In Fig.\ref{fig:FID}(b), it can be seen that the magnetic anisotropy of singlet yield mainly comes from the coherent phase changes $\phi_{mn}^{\hat{B}}(t) $ induced by the Earth's magnetic field, as in the interference-based quantum metrology. 
The analogy between a quantum interferometer and a chemical compass (a complete list of analogies is included in \cite{SI})
 now allows us to clearly identify the role of coherence in chemical magnetoreception in a quantitative manner.

{\it Quantum coherence and compass sensitivity.---} The function of a chemical compass starting from the state $\rho_0$ can be characterized by the magnetic anisotropy of the singlet yield defined as follows
\be
\mbox{D}_{\mbox{s}}(\mathcal{H}_0,\rho_0)=\max_{\theta,\phi} \mbox{Y}_{\mbox{s}}(\mathcal{H}_0,\rho_0,\theta,\phi)-\min_{\theta,\phi} \mbox{Y}_{\mbox{s}}(\mathcal{H}_0,\rho_0,\theta,\phi)\label{eq:magani}
\ee
Henceforth we will call $\mbox{D}_{\mbox{s}}(\mathcal{H}_0,\rho_0)$ as the magnetic sensitivity of a chemical compass. The coherent part of the system state $\rho_0$ is denoted as
\be
\mathcal{GC}(\mathcal{H}_0,\rho_0) = \sum_{m\neq n} r_{mn} \ket{\nu_m}\bra{\nu_n}. 
\ee
To quantify the role of coherence in chemical magnetoreception, following the inspiration from quantifying the role of coherence in quantum interferometric metrology (see Eq.(\ref{eq:mcqm})), we introduce the following measure of global electron-nuclear quantum coherence for a chemical compass with a given hyperfine interaction Hamiltonian $\mathcal{H}_0$ and a system state $\rho_0$ as follows
\be
\mathcal{C}(\mathcal{H}_0,\rho_0)=\left\vert\mbox{Y}_{\mbox{s}}(\mathcal{H}_0,\mathcal{GC}(\mathcal{H}_0,\rho_0))\right\vert_{b=0}\label{eq:gcm}
\ee
which represents the contribution of coherence to the singlet yield in the absence of the Earth's magnetic field.

\begin{figure}[t]
\begin{center}
\hspace{-0.32cm}
\includegraphics[width=4.47cm]{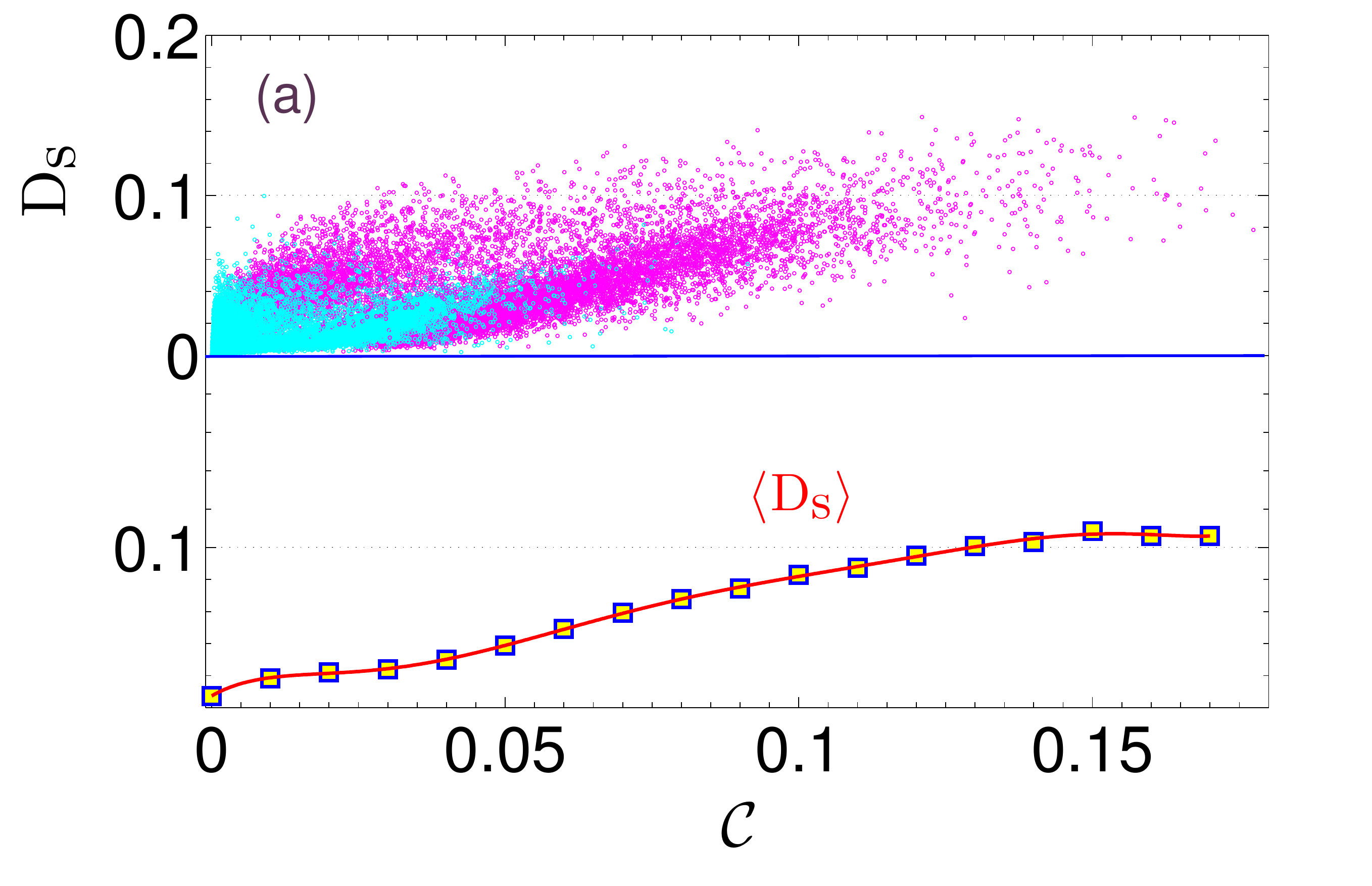}
\hspace{-0.25cm}
\includegraphics[width=4.4cm]{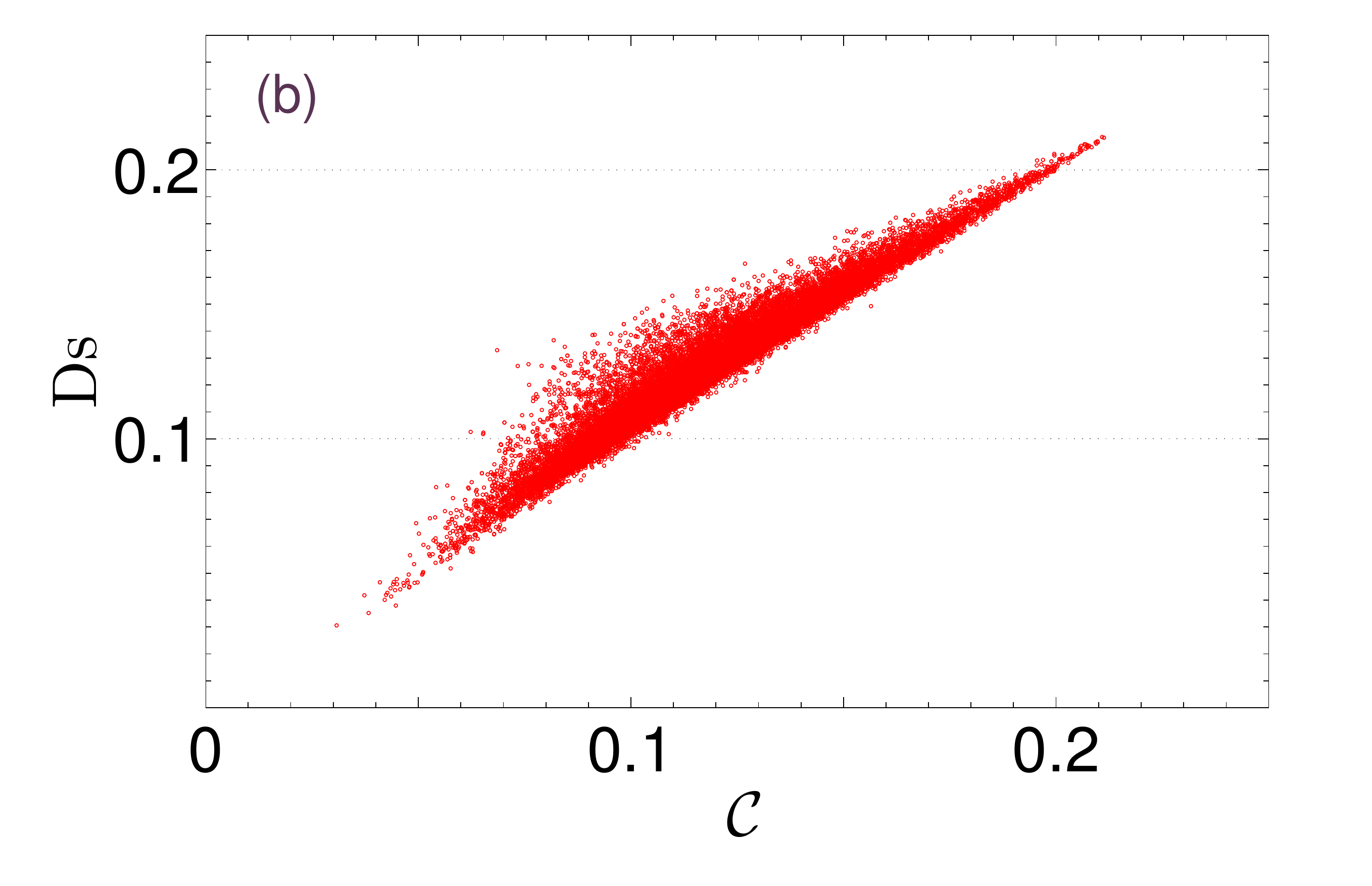}
\end{center}
\caption{(a) The magnetic field sensitivity $\mbox{D}_{\mbox{s}}(\mathcal{H}_0,\rho_0)$ as defined in Eq.(\ref{eq:magani}) (upper) and the average value $\langle \mbox{D}_{\mbox{s}} \rangle $ (lower) as a function of coherence $\mathcal{C}(\mathcal{H}_0,\rho_0)$ given by Eq.(\ref{eq:gcm}) for $7\times 10^4$ random hyperfine configurations with in total 5-6 nuclei initially in the depolarized state. The violet dots represent these configurations of reference-and-probe type molecules (one radical is free from hyperfine coupling). (b) The magnetic field sensitivity $\mbox{D}_{\mbox{s}}(\mathcal{H}_{\mbox{f}},\rho_0)$ as a function of coherence $\mathcal{C}(\mathcal{H}_{\mbox{f}},\rho_0)$ for a radical pair (in which one radical contains three $^1$H nuclei and two $^{14}$N nuclei as FADH$\cdot$) starting from $2\times 10^4$ initial states, each has the singlet born radical pair and random nuclear spin polarizations. In both panels, the reaction rates are $k_S=k_T=0.5 \mu s^{-1}$, and the Earth's magnetic field is $b=50\mu T$.}\label{fig:CRDS}
\end{figure}

To demonstrate the connection between global coherence and the magnetic sensitivity in a chemical compass for general molecules, we randomly sample a large number of hyperfine configurations \cite{RS} and plot in Fig.\ref{fig:CRDS}(a) the compass sensitivity $\mbox{D}_{\mbox{s}}(\mathcal{H}_0,\rho_0)$ as a function of coherence $\mathcal{C}(\mathcal{H}_0,\rho_0)$ and , where $\rho_0=\ketbra{\mathcal{S}}{\mathcal{S}}\bigotimes_{k_j} (\mathbb{I}_{d_{k_j}}/d_{k_j})$. The result shows that typically the global coherence is a {\it resource} for the function of a chemical compass, namely the larger coherence the better magnetic sensitivity. We find that global coherence makes more dominant contribution to the compass sensitivity as compared with local electronic coherence \cite{SI}. The perspective that coherence is a resource for chemical magnetoreception offers a unique guide towards various design principles of radical pair molecules. For example, one can see that the reference-and-probe type of radical pairs (namely one radical of which is free from hyperfine coupling) \cite{Ritz10,Maeda08} which tends to result in the highest sensitivity usually embody larger coherence. Considering the explicit example of the radical pair inspired by the flavin adenine dinucleotide in cryptochromes, we plot in Fig.\ref{fig:CRDS}(b) the magnetic sensitivity as a function of coherence when varying the nuclear spin polarization. The same feature is observed, namely the larger the global coherence is, the better the magnetic sensitivity. We remark that the nuclear spins are usually depolarized at ambient temperature. We assume here a larger state space to facilitate the explicit demonstration of how coherence (that changes with the nuclear polarization) can directly affect the function of chemical magnetoreception. In experiments, this may be achieved by chemically induced dynamic nuclear polarization or with the assistance of quantum control via, e.g. nitrogen-vacancy centers in diamond \cite{Cai12_njp,Paz13,Cai13}. In contrast, the prominent connection is absent between the compass sensitivity and local electronic coherence \cite{SI}. We also stress that the above measure of coherence in Eq.(\ref{eq:gcm}) is determined by the hyperfine interaction Hamiltonian $\mathcal{H}_0$ (without an external magnetic field) and the joint system state $\rho_0$, yet it well predicts the property (namely the magnetic sensitivity) of the dynamics of a chemical compass when changing the magnetic field direction.  

{\it A unified picture of decoherence effects.---} Following the present insight that quantum coherence of the global electron-nuclear state is a resource for chemical magnetoreception, it is possible to study the effects of different decoherence models on the functioning of a chemical compass \cite{Gauger11,Cai12,Band12} in a unified picture. More specifically, we can study how decoherence will destroy the global electron-nuclear quantum coherence in a chemical compass, as characterized by Eq.(\ref{eq:gcm}), and thereby deteriorate its magnetic anisotropy of reaction yield. Since a chemical compass shall work under ambient conditions, the noise from the environment of the core system (i.e. the radical pair and the surrounding nuclear spins) will inevitably affect its function. The noise effects vary for different decoherence models, which can be described by the Lindblad type quantum master equation as \cite{Gauger11}
\be
\frac{\mbox{d}}{\mbox{dt}}\rho=-i[\mathcal{H},\rho]-\frac{k_S}{2}[Q_S,\rho]_{+}-\frac{k_T}{2}[Q_T,\rho]_{+}+\mathcal{L}(\rho)
\ee
with $
\mathcal{L}(\rho)=\sum_k\xi_k\blb{ \mathcal{L}_k \rho \mathcal{L}_k^{\dagger}-\frac{1}{2}\bla{\mathcal{L}_k^{\dagger} \mathcal{L}_k \rho-\rho \mathcal{L}_k^{\dagger} \mathcal{L}_k}}$, where $\mathcal{H}=\mathcal{H}_0+\mathrm{\hat{B}}(\theta,\phi)$ is the total Hamiltonian including the hyperfine interactions $\mathcal{H}_0$ and the external magnetic field $\mathrm{\hat{B}}(\theta,\phi)$, $Q_S$ and $Q_T$ are the projectors into the singlet and triplet subspace of the radical pair state individually, and $[x,y]_{+}=xy+yx$. The above master equation is based on the Haberkorn approach \cite{Steiner89} by adding the dissipator $\mathcal{L}(\rho)$ which represent the environmental noise. We remark that it is not trace preserving as it is restricted in the subspace of the active radical pair state, and the singlet yield can then be calculated as $\mbox{Y}_{\mbox{s}}=k_S\int \bra{\mathcal{S}} \rho(t)\ket{\mathcal{S}}dt$, which is equivalent to Eq.(\ref{singletyield}) in the absence of decoherence (see Ref.\cite{Hore10}). We consider three typical classes of environmental noise (that are independent on the magnetic field by themselves \cite{Gauger11}), namely the local dephasing model $\mathcal{L}_{I}=\{\sigma_z^{D},\sigma_z^{A}\}$ with $\sigma_z=\ketbra{\alpha}{\alpha}-\ketbra{\beta}{\beta}$, the spin relaxation $\mathcal{L}_{II}=\{\sigma_{\pm}^{D} ,\sigma_{\pm}^{A}\}$ with $\sigma_{+}=\ketbra{\alpha}{\beta}$ and $\sigma_{-}=\ketbra{\beta}{\alpha}$, and the singlet-triplet dephasing model \cite{Shu91} $\mathcal{L}_{III}=\{\sigma_{ST}^{DA}\}$ with $\sigma_{ST}=2\ket{\mathcal{S}}\bra{\mathcal{S}}-\mathbb{I}$, where the superscript $D$ and $A$ represent two radicals in a pair. These decoherence models are most relevant and were included to fit the experiment data \cite{Maeda12} but behave very different from each other in the sense that they destroy varying aspects of radical pair coherence: they will respectively lead to the decay of local electron spin coherence, of electron spin longitudinal component, and of singlet-triplet coherence at a rate given by $\xi$.

\begin{figure}[t]
\begin{center}
\hspace{-0.36cm}
\includegraphics[width=4.56cm]{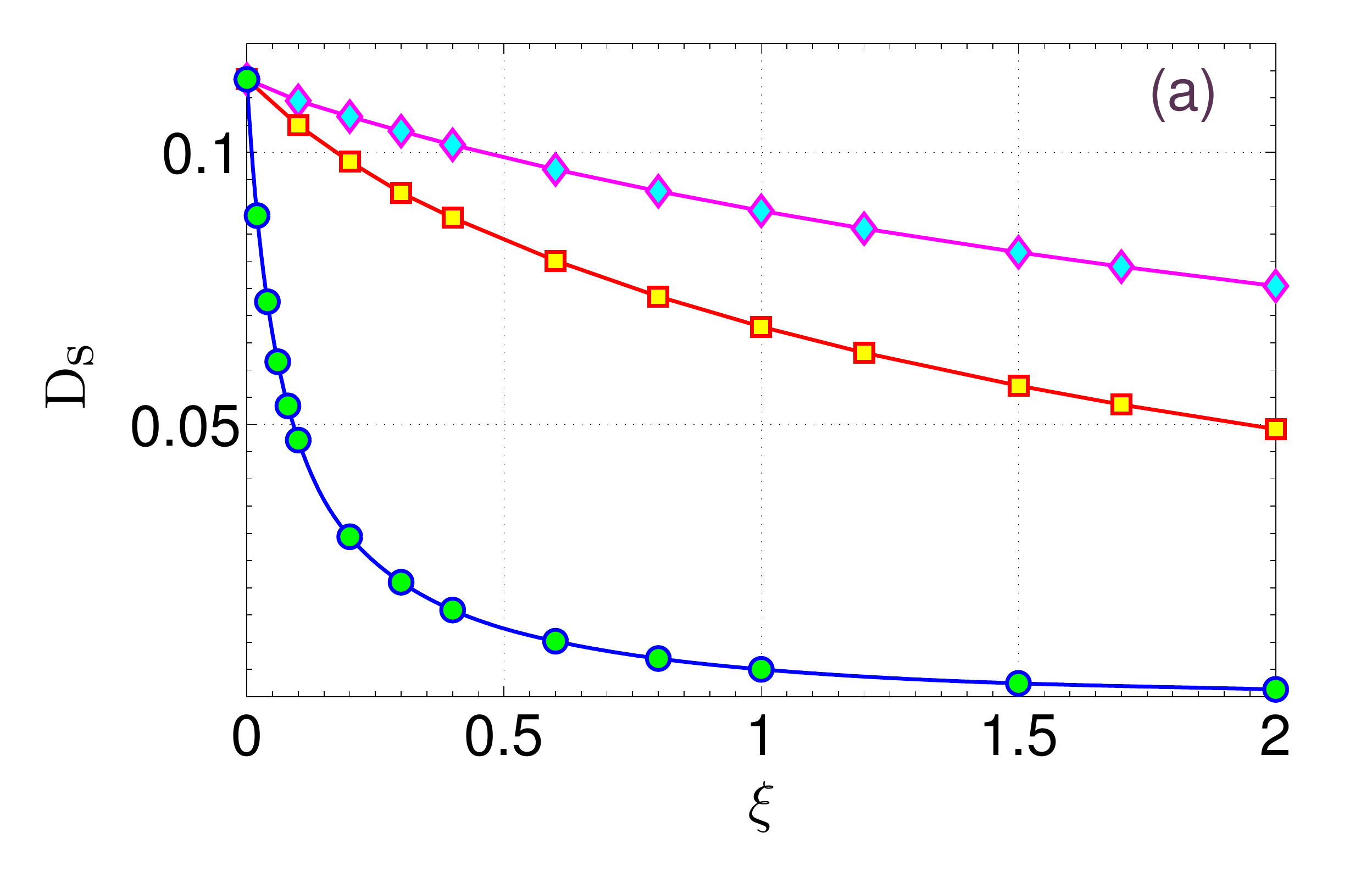}
\hspace{-0.46cm}
\includegraphics[width=4.58cm]{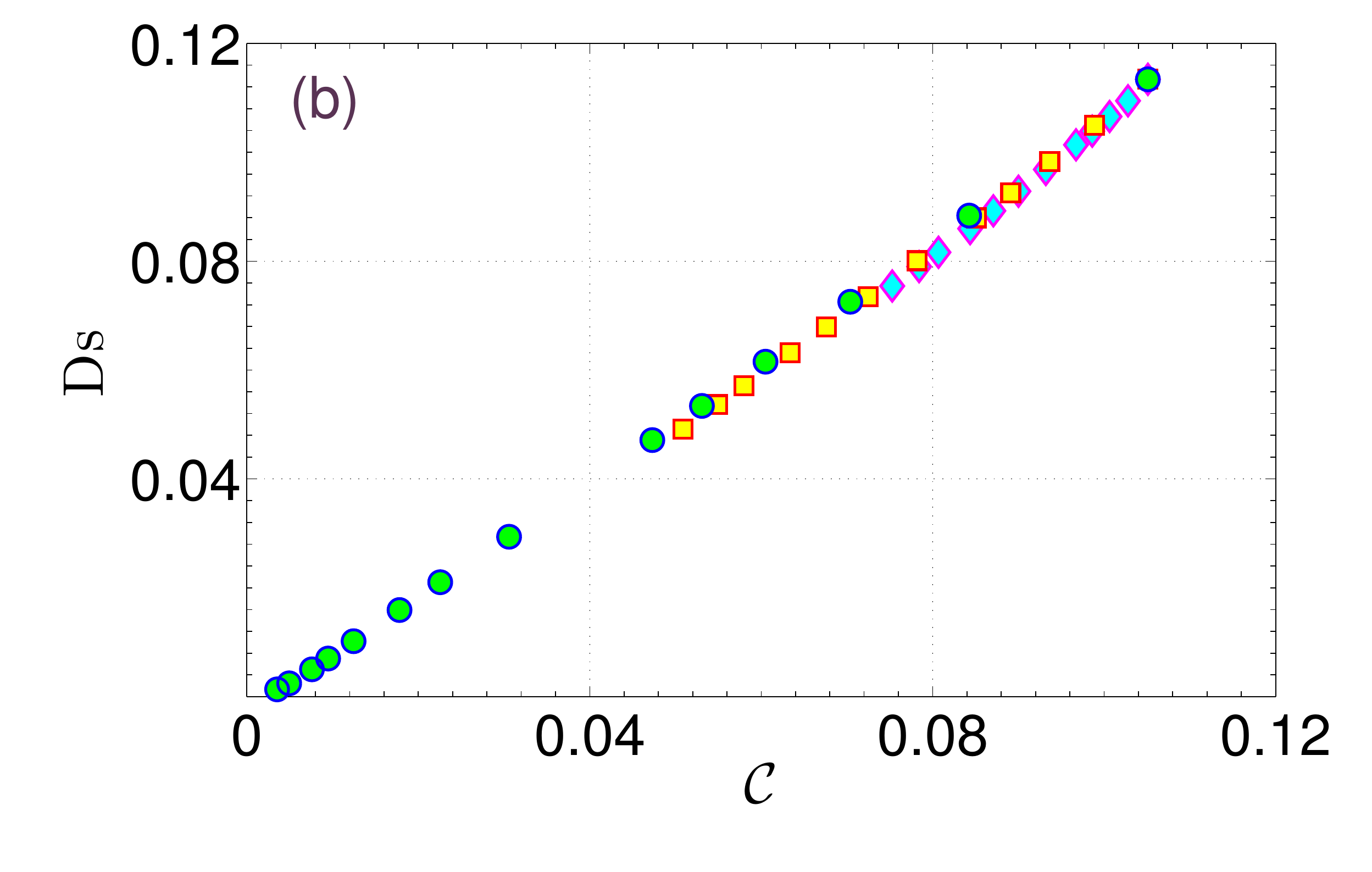}
\end{center}
\caption{The noise effect on the magnetic sensitivity of a radical pair (in which one radical contains three $^1$H nuclei and two $^{14}$N nuclei as FADH$\cdot$) for the local dephasing (red, square), the electron spin relaxation (blue, circle), and the singlet-triplet dephasing (violet, diamond). (a) The magnetic sensitivity $\mbox{D}_{\mbox{s}}(\mathcal{H}_0,\rho_0)$  as a function of the decoherence rate $\xi$ (in the unit of $\mu s^{-1}$). (b) The magnetic sensitivity $\mbox{D}_{\mbox{s}}(\mathcal{H}_0,\rho_0)$ as a function of the coherence $\mathcal{C}(\mathcal{H}_0,\rho_0)$. In both panels, the reaction rates are $k_S=k_T=0.5 \mu s^{-1}$, and the Earth's magnetic field is $b=50\mu T$.}\label{fig:DEC}
\end{figure}

In Fig.\ref{fig:DEC}(a), one can see that the effects of these noise models on the magnetic sensitivity of the radical pair are quantitatively very different as a function of the same decoherence rate $\xi$. Such a fact is natural but gives no general insight into the problem how and why different types of noise would affect chemical magnetoreception to different extent. Instead, we calculate the global coherence $\mathcal{C}$ (see Eq.\ref{eq:gcm}) under the influence of environmental noise and find a universal relation between the decoherence effects on the magnetic sensitivity and the coherence, which holds for different noise models studied here, see Fig.\ref{fig:DEC}(b). We have checked that such a universal relation also holds for mutations of the radical pair through (partially) deuteration, which will be interesting to study in spin chemistry experiment.  This supports the observation that the global coherence is an appropriate concept to quantify the role of coherence in chemical magnetoreception. This is demonstrated by a study of different noise sources whose effect on the magnetic sensitivity can be predicted from their effect on the global coherence.

{\it Discussion and outlook.---} We have introduced the viewpoint of chemical magnetoreception as a quantum interferometer. This perspective allows us to reveal the direct connection between the global electron-nuclear spin coherence and the magnetic sensitivity of chemical magnetoreception for general molecules, it thus evidences coherence as a resource in the chemical compass model of avian magnetoreception in a similar way as a coherence-based quantum device. The verification of the present observation is conceivable either in spin chemistry experiments or by  quantum simulation with well controllable systems, for example nitrogen-vacancy defects in diamond \cite{Dolde13, Cai13}, to emulate the radical pair dynamics. We remark that coherent manipulations and readout of electron or nuclear spins in diamond have been well developed in experiments. The concept of global coherence offers a unified perspective to predict the magnetic sensitivity of a chemical compass. Therefore, it is possible to use it to advance our understanding of the design principles of a chemical compass, namely to exhibit a better sensitivity the molecule shall embody a larger global coherence. This will facilitate the construction of an artificial chemical compass which is sensitive to the weak geomagnetic field at ambient temperature \cite{Maeda08}, the demonstration of which would serve as an important intermediate step to understand how nature might be able to design such a chemical compass. We expect that the present ideas may help to further transfer the concepts and methods developed in quantum information to the field of spin chemistry, and gain new insights into the other coherent phenomena in spin chemistry, such as low field effect, coherence transfer in spin correlated radical pairs \cite{Wasie11} and other variants of radical pair model of magnetoreception \cite{Lovett12,Carrillo13}. 

{\it Acknowledgements.---} We are grateful for the valuable discussions with Susana F. Huelga, Till Biskup, Ulrich E. Steiner, Erik Gauger, Peter Hore and Ilia Solov'yov. The work was supported by the Alexander von Humboldt Foundation, EU Integrating Projects Q-ESSENCE and SIQS, EU STREP PAPETS, ERC Synergy grant BioQ. J.-M.C was supported also by a Marie-Curie Intra-European Fellowship (FP7). Computations were performed on the bwGRiD.\\


\begin{thebibliography}{99}
\providecommand{\url}[1]{\texttt{#1}}
\providecommand{\urlprefix}{URL }
\providecommand{\eprint}[2][]{\url{#2}}


\bibitem{Engel07} G. S. Engel, T. R. Calhoun, E. L. Read, T.-K. Ahn, T. Manal, Y.-C. Cheng, R. E. Blankenship, G. R. Fleming, Nature {\bf 446}, 782-786 (2007).

\bibitem{Collini10} E. Collini, C. Y. Wong, K. E. Wilk, P. M. G. Curmi, P. Brumer, G. D. Scholes, Nature 463, 644-647 (2010).

\bibitem{LHCT1} M. Mohseni, P. Rebentrost, S. Lloyd, and A. Aspuru-Guzik, J. Chem. Phys. \textbf{129}, 174106 (2008).

\bibitem{Plenio08} M. B. Plenio and S. F. Huelga, New J. Phys. \textbf{10}, 113019 (2008).

\bibitem{Scho11} G. D. Scholes, G. R. Fleming, A. Olaya-Castro, and R. van Grondelle, Nature Chemistry {\bf 3}, 763-774 (2011).

\bibitem{Chin13} A. W. Chin, J. Prior, R. Rosenbach, F. Caycedo-Soler, S. F. Huelga, and M. B. Plenio, Nature Physics. {\bf 9}, 113-118 (2013).

\bibitem{SME1} L. Turin, J. Theor. Biol. \textbf{216}(3), 367 (2002).

\bibitem{SME2} J. C. Brookes, F. Hartoutsiou, A. P. Horsfield, and A. M. Stoneham, Phys. Rev. Lett. \textbf{98}, 038101 (2007).

\bibitem{SME3} I. A. Solov'yov, P.-Y. Chang, and K. Schulten, Phys. Chem. Chem. Phys. {\bf 14}, 13861-13871 (2012) .

\bibitem{Sch78} K. Schulten, C. E. Swenberg, and A. Weller, Z. Phys. Chem \textbf{NF111}, 1-5 (1978).

\bibitem{Ritz00} T. Ritz, S. Adem, and K. Schulten, Biophys. J. {\bf 78}, 707-718 (2000).

\bibitem{Sol07} Ilia A. Solov'yov, D. Chandler, and K. Schulten, Biophys. J \textbf{92}, 2711 (2007).

\bibitem{Rodgers09} C. T. Rodgers and P. J. Hore, Proc. Natl. Acad. Sci. U.S.A. {\bf 106}, 353 (2009).

\bibitem{Ritz09} T. Ritz, R. Wiltschko, P. J. Hore, C. T. Rodgers, K. Stapput, P. Thalau, C. R. Timmel, W. Wiltschko, Biophys. J. {\bf 96}, 3451 (2009).

\bibitem{Ritz10} T. Ritz, M. Ahmad, H. Mouritsen, R. Wiltschko, W. Wiltschko, J. Royal. Soc. Interface {\bf 7}, S135-146 (2010).

\bibitem{Maeda08} K. Maeda, K. B. Henbest, F. Cintolesi, I. Kuprov, C. T. Rodgers, P. A. Liddell, D. Gust, C. R. Timmel, P. J. Hore, Nature {\bf 453}, 387-390 (2008).

\bibitem{Ritz11} T. Ritz, Procedia Chemistry {\bf 3}, 262 (2011). 

\bibitem{Hore10} J.~A. Jones, P. J. Hore, Chem. Phys. Lett. 488, 90 (2010).

\bibitem{Kom09} I. K. Kominis, Phys. Rev. E {\bf 80}, 056115 (2009); I. K. Kominis, Phys. Rev. E {\bf 83}, 056118 (2011).

 \bibitem{Cai10} J.-M. Cai, G.~G. Guerreschi, and H.~J. Briegel, Phys. Rev. Lett. {\bf 104}, 220502 (2010).

\bibitem{Gauger11} E. M. Gauger, E. Rieper, J. J. L. Morton, S. C. Benjamin, and V. Vedral, Phys. Rev. Lett. {\bf 106}, 040503 (2011).

\bibitem{Sun12} C. Y. Cai, Qing Ai, H. T. Quan, C. P. Sun, Phys. Rev. A {\bf 85}, 022315 (2012).

\bibitem{Cai12} J.-M. Cai, F. Caruso, and M. B. Plenio, Phys. Rev. A {\bf 85}, 040304 (2012).

\bibitem{Band12} J. N. Bandyopadhyay, T. Paterek, and D. Kaszlikowski,  Phys. Rev. Lett. {\bf 109}, 110502 (2012); E. M. Gauger, S. C. Benjamin, Phys. Rev. Lett. {\bf 110}, 178901 (2013); J. N. Bandyopadhyay, T. Paterek, and D. Kaszlikowski, Phys. Rev. Lett. {\bf 110}, 178902 (2013).

\bibitem{Hogben12} H. J. Hogben, T. Biskup, and P. J. Hore, Phys. Rev. Lett. {\bf 109}, 220501 (2012).

%

\bibitem{Guer2013} G. G. Guerreschi, M. Tiersch, U. Steiner, H. J. Briegel, Chem. Lett. {\bf 572}, 106 (2013).

\bibitem{Joh05} S. Johnsen and K. J. Lohmann, Nature Rev. Neurosci. {\bf 6}, 703-712 (2005).

\bibitem{Werner77} H.‐J. Werner, Z. Schulten, and K. Schulten, J. Chem. Phys. {\bf 67}, 646 (1977).

\bibitem{Mour04} H. Mouritsen, U. Janssen-Bienhold, M. Liedvogel, G. Feenders, J. Stalleicken, P. Dirks, R. Weiler, Proc. Natl Acad. Sci. USA {\bf 101}, 14294-14299 (2004). 

\bibitem{Weaver00nature}  J. C. Weaver, T. E. Vaughan and R. D. Astumian, Nature {\bf 405}, 707-709 (2000). 

\bibitem{Hore03}  F. Cintolesi, T. Ritz, C. W. M. Kay, C. R. Timmel and P. J. Hore, Chem. Phys. {\bf 294}, 385-399 (2003). 

\bibitem{Solo07} I. A. Solov’yov, D. E. Chandler and K. Schulten, Biophys. J. {\bf 92}, 2711-2726 (2007).

\bibitem{Ahmad07} M. Ahmad, P. Galland, T. Ritz, R. Wiltschko and W.  Wiltschko, Planta {\bf 225}, 615-624 (2007). 

\bibitem{Zap09} M. Zapka, D. Heyers, C. M. Hein, S. Engels, N.-L. Schneider, J. Hans, S.  Weiler, D. Dreyer, D. Kishkinev, J. M. Wild, H.  Mouritsen,  Nature \textbf{461}, 1274-1277 (2009).

\bibitem{Maeda12} K. Maeda, A. J. Robinson, K. B. Henbest, H. J. Hogben, T. Biskup, M. Ahmad, E. Schleicher, S. Weber, C. R. Timmel, and P. J. Hore, Proc. Natl. Acad. Sci. U.S.A. {\bf 109}, 4774 (2012). 

\bibitem{Sol10} I. A. Solov'yov, H. Mouritsen, K. Schulten, Biophy. J. {\bf 99}, 40-49 (2010).

\bibitem{Sol12} I. A. Solov'yov, T. Domratcheva, A. R. M. Shahi, and K. Schulten, Journal of the American Chemical Society. {\bf 134}, 8046-1805 (2012). 

\bibitem{Lied10} M. Liedvogel, H. Mouritsen, J. Roy. Soc. Interface. {\bf 7}, S147-S162  (2010).

\bibitem{Phil92} J. B Phillips, and S. C. Borland, Nature {\bf 359}, 142-144 (1992).

\bibitem{WiltsRev05} W. Wiltschko, R. Wiltschko, J. Comp. Physiol. A. {\bf 191}, 675-693 (2005).

\bibitem{Ritz04nature} T. Ritz, P. Thalau, J. B. Phillips, R. Wiltschko and W. Wiltschko, Nature {\bf 429}, 177-180 (2004). 

\bibitem{Steiner89} U. E. Steiner, T. Ulrich, Chem. Rev. {\bf 89}, 51 (1989).

\bibitem{Wine94} D. J. Wineland, J. J. Bollinger, W. M. Itano, and  D. J. Heinzen, Phys. Rev. A {\bf 50}, 67-88 (1994).


\bibitem{Budker07} D. Budker, and M. Romalis, Nature Physics {\bf 3}, 227 - 234 (2007).



\bibitem{Caves07} A. Shaji and C. M. Caves, Phys. Rev. A {\bf 76}, 032111 (2007).

\bibitem{Efi08} O. Efimova and P. J. Hore, Biophys. J., \textbf{94}, 1565-1574 (2008).

\bibitem{SI} See \href{http://link.aps.org/supplemental/10.1103/PhysRevLett.111.230503}{Supplemental Material} for the calculation details and more discussions.

\bibitem{RS} The radical pair that is responsible for the magnetic sense of birds has not been conclusively determined. In the present work, we thus explore the essential features of typical radical pair molecules (which in total consist of more than three up to seven nuclei with random hyperfine interaction strengths \cite{SI}) rather than restricting to the radical pairs with very specific hyperfine configurations, particularly we go beyond the simplified model with only one single nucleus \cite{Gauger11,Cai12,Hogben12,Band12}. Besides the typical behaviors of general radical pair molecules, we also take the radical pair inspired by the flavin adenine dinucleotide (FADH$\cdot$) formed photochemically in cryptochromes as an explicit example to further elaborate our results.

\bibitem{Cai12_njp} J.-M. Cai, F. Jelezko, M. B. Plenio, A. Retzker, New J. Phys. {\bf 15}, 013020 (2013).

\bibitem{Cai13} J.-M. Cai, A. Retzker, F. Jelezko, M. B. Plenio, Nature Physics {\bf 9}, 168 (2013).

\bibitem{Paz13} P. London,  J. Scheuer, J.-M. Cai, I. Schwarz, A. Retzker, M.B. Plenio, M. Katagiri, T. Teraji, S. Koizumi, J. Isoya, R. Fischer, L. P. McGuinness, B. Naydenov, and F. Jelezko, Phys. Rev. Lett. {\bf 111}, 067601 (2013).

\bibitem{Shu91} A. I. Shushin, Chem. Phys. Lett. {\bf 181}, 274-278 (1991).

\bibitem{Dolde13} F. Dolde,	I. Jakobi, B. Naydenov, N. Zhao, S. Pezzagna, C. Trautmann, J. Meijer,	P. Neumann, F. Jelezko, J. Wrachtrup, Nature Physics {\bf 9}, 139 (2013). 

\bibitem{Wasie11} M. T. Colvin, A. Butler Ricks, A. M. Scott, A. L. Smeigh, R. Carmieli, T. Miura, and M. R. Wasielewski, J. Am. Chem. Soc. {\bf 133}, 1240-1243 (2011).

\bibitem{Lovett12} B. W. Lovett, et al,  Biophys. J., {\bf 102}, 961 (2012).

\bibitem{Carrillo13} A. Carrillo, M. F. Cornelio, M. C. de Oliveira,  arXiv:1304.3452.

\end{thebibliography}
\end{document}